\magnification=\magstep1



\catcode`\X=12\catcode`\@=11

\def\n@wcount{\alloc@0\count\countdef\insc@unt}
\def\n@wwrite{\alloc@7\write\chardef\sixt@@n}
\def\n@wread{\alloc@6\read\chardef\sixt@@n}
\def\r@s@t{\relax}\def\v@idline{\par}\def\@mputate#1/{#1}
\def\l@c@l#1X{\firstpart.#1}\def\gl@b@l#1X{#1}\def\t@d@l#1X{{}}

\def\crossrefs#1{\ifx\all#1\let\tr@ce=\all\else\def\tr@ce{#1,}\fi
   \n@wwrite\cit@tionsout\openout\cit@tionsout=\jobname.cit 
   \write\cit@tionsout{\tr@ce}\expandafter\setfl@gs\tr@ce,}
\def\setfl@gs#1,{\def\@{#1}\ifx\@\empty\let\next=\relax
   \else\let\next=\setfl@gs\expandafter\xdef
   \csname#1tr@cetrue\endcsname{}\fi\next}
\def\m@ketag#1#2{\expandafter\n@wcount\csname#2tagno\endcsname
     \csname#2tagno\endcsname=0\let\tail=\all\xdef\all{\tail#2,}
   \ifx#1\l@c@l\let\tail=\r@s@t\xdef\r@s@t{\csname#2tagno\endcsname=0\tail}\fi
   \expandafter\gdef\csname#2cite\endcsname##1{\expandafter
     \ifx\csname#2tag##1\endcsname\relax?\else\csname#2tag##1\endcsname\fi
     \expandafter\ifx\csname#2tr@cetrue\endcsname\relax\else
     \write\cit@tionsout{#2tag ##1 cited on page \folio.}\fi}
   \expandafter\gdef\csname#2page\endcsname##1{\expandafter
     \ifx\csname#2page##1\endcsname\relax?\else\csname#2page##1\endcsname\fi
     \expandafter\ifx\csname#2tr@cetrue\endcsname\relax\else
     \write\cit@tionsout{#2tag ##1 cited on page \folio.}\fi}
   \expandafter\gdef\csname#2tag\endcsname##1{\expandafter
      \ifx\csname#2check##1\endcsname\relax
      \expandafter\xdef\csname#2check##1\endcsname{}%
      \else\immediate\write16{Warning: #2tag ##1 used more than once.}\fi
      \multit@g{#1}{#2}##1/X%
      \write\t@gsout{#2tag ##1 assigned number \csname#2tag##1\endcsname\space
      on page \number\count0.}%
   \csname#2tag##1\endcsname}}
\def\multit@g#1#2#3/#4X{\def\t@mp{#4}\ifx\t@mp\empty%
      \global\advance\csname#2tagno\endcsname by 1 
      \expandafter\xdef\csname#2tag#3\endcsname
      {#1\number\csname#2tagno\endcsnameX}%
   \else\expandafter\ifx\csname#2last#3\endcsname\relax
      \expandafter\n@wcount\csname#2last#3\endcsname
      \global\advance\csname#2tagno\endcsname by 1 
      \expandafter\xdef\csname#2tag#3\endcsname
      {#1\number\csname#2tagno\endcsnameX}%
      \write\t@gsout{#2tag #3 assigned number \csname#2tag#3\endcsname\space
      on page \number\count0.}\fi
   \global\advance\csname#2last#3\endcsname by 1
   \def\t@mp{\expandafter\xdef\csname#2tag#3/}%
   \expandafter\t@mp\@mputate#4\endcsname
   {\csname#2tag#3\endcsname\lastpart{\csname#2last#3\endcsname}}\fi}
\def\t@gs#1{\def\all{}\m@ketag#1e\m@ketag#1s\m@ketag\t@d@l p
   \m@ketag\gl@b@l r \n@wread\t@gsin
   \openin\t@gsin=\jobname.tgs \re@der \closein\t@gsin
   \n@wwrite\t@gsout\openout\t@gsout=\jobname.tgs }
\outer\def\localtags{\t@gs\l@c@l}
\outer\def\globaltags{\t@gs\gl@b@l}
\outer\def\newlocaltag#1{\m@ketag\l@c@l{#1}}
\outer\def\newglobaltag#1{\m@ketag\gl@b@l{#1}}

\newif\ifpr@ 
\def\m@kecs #1tag #2 assigned number #3 on page #4.%
   {\expandafter\gdef\csname#1tag#2\endcsname{#3}
   \expandafter\gdef\csname#1page#2\endcsname{#4}
   \ifpr@\expandafter\xdef\csname#1check#2\endcsname{}\fi}
\def\re@der{\ifeof\t@gsin\let\next=\relax\else
   \read\t@gsin to\t@gline\ifx\t@gline\v@idline\else
   \expandafter\m@kecs \t@gline\fi\let \next=\re@der\fi\next}
\def\pretags#1{\pr@true\pret@gs#1,,}
\def\pret@gs#1,{\def\@{#1}\ifx\@\empty\let\n@xtfile=\relax
   \else\let\n@xtfile=\pret@gs \openin\t@gsin=#1.tgs \message{#1} \re@der 
   \closein\t@gsin\fi \n@xtfile}

\newcount\sectno\sectno=0\newcount\subsectno\subsectno=0
\newif\ifultr@local \def\ultralocal{\ultr@localtrue}
\def\firstpart{\number\sectno}
\def\lastpart#1{\ifcase#1 \or a\or b\or c\or d\or e\or f\or g\or h\or 
   i\or k\or l\or m\or n\or o\or p\or q\or r\or s\or t\or u\or v\or w\or 
   x\or y\or z \fi}

\def\resetall{\global\advance\sectno by 1\subsectno=0
   \gdef\firstpart{\number\sectno}\r@s@t}
\def\resetsub{\global\advance\subsectno by 1
   \gdef\firstpart{\number\sectno.\number\subsectno}\r@s@t}
\def\newsection#1\par{\resetall\vskip0pt plus.3\vsize\penalty-250
   \vskip0pt plus-.3\vsize\bigskip\bigskip
   \message{#1}\leftline{\bf#1}\nobreak\bigskip}
\def\subsection#1\par{\ifultr@local\resetsub\fi
   \vskip0pt plus.2\vsize\penalty-250\vskip0pt plus-.2\vsize
   \bigskip\smallskip\message{#1}\leftline{\bf#1}\nobreak\medskip}

\def\t@gsoff#1,{\def\@{#1}\ifx\@\empty\let\next=\relax\else\let\next=\t@gsoff
   \def\@@{p}\ifx\@\@@\else
   \expandafter\gdef\csname#1cite\endcsname{\relax}
   \expandafter\gdef\csname#1page\endcsname##1{?}
   \expandafter\gdef\csname#1tag\endcsname{\relax}\fi\fi\next}
\def\verbatimtags{\ifx\all\relax\else\expandafter\t@gsoff\all,\fi}

\def\(#1){\edef\dot@g{\ifmmode\ifinner(\hbox{\rm \noexpand\etag{#1}})
   \else\noexpand\eqno(\hbox{\noexpand\etag{#1}})\fi
   \else(\noexpand\ecite{#1})\fi}\dot@g}

\newif\ifbr@ck
\def\eat#1{}
\def\[#1]{\br@cktrue[\br@cket#1'X]}
\def\br@cket#1'#2X{\def\temp{#2}\ifx\temp\empty\let\next\eat
   \else\let\next\br@cket\fi
   \ifbr@ck\br@ckfalse\br@ck@t#1,X\else\br@cktrue#1\fi\next#2X}
\def\br@ck@t#1,#2X{\def\temp{#2}\ifx\temp\empty\let\neext\eat
   \else\let\neext\br@ck@t\def\temp{,}\fi
   \def\teemp{#1}\ifx\teemp\empty\else\rcite{#1}\fi\temp\neext#2X}
\def\resetbr@cket{\gdef\[##1]{[\rtag{##1}]}}
\def\references{\resetbr@cket\newsection References\par}

\newtoks\symb@ls\newtoks\s@mb@ls\newtoks\p@gelist\n@wcount\ftn@mber
    \ftn@mber=1\newif\ifftn@mbers\ftn@mbersfalse\newif\ifbyp@ge\byp@gefalse
\def\defm@rk{\ifftn@mbers\n@mberm@rk\else\symb@lm@rk\fi}
\def\n@mberm@rk{\xdef\m@rk{{\the\ftn@mber}}%
    \global\advance\ftn@mber by 1 }
\def\rot@te#1{\let\temp=#1\global#1=\expandafter\r@t@te\the\temp,X}
\def\r@t@te#1,#2X{{#2#1}\xdef\m@rk{{#1}}}
\def\b@@st#1{{$^{#1}$}}\def\str@p#1{#1}
\def\symb@lm@rk{\ifbyp@ge\rot@te\p@gelist\ifnum\expandafter\str@p\m@rk=1 
    \s@mb@ls=\symb@ls\fi\write\f@nsout{\number\count0}\fi \rot@te\s@mb@ls}
\def\byp@ge{\byp@getrue\n@wwrite\f@nsin\openin\f@nsin=\jobname.fns 
    \n@wcount\currentp@ge\currentp@ge=0\p@gelist={0}
    \re@dfns\closein\f@nsin\rot@te\p@gelist
    \n@wread\f@nsout\openout\f@nsout=\jobname.fns }
\def\m@kelist#1X#2{{#1,#2}}
\def\re@dfns{\ifeof\f@nsin\let\next=\relax\else\read\f@nsin to \f@nline
    \ifx\f@nline\v@idline\else\let\t@mplist=\p@gelist
    \ifnum\currentp@ge=\f@nline
    \global\p@gelist=\expandafter\m@kelist\the\t@mplistX0
    \else\currentp@ge=\f@nline
    \global\p@gelist=\expandafter\m@kelist\the\t@mplistX1\fi\fi
    \let\next=\re@dfns\fi\next}
\def\symbols#1{\symb@ls={#1}\s@mb@ls=\symb@ls} 
\def\bigsymbol{\textstyle}
\symbols{\bigsymbol\ast,\dagger,\ddagger,\sharp,\flat,\natural,\star}
\def\ftnumbers{\ftn@mberstrue} \def\ftsymbols{\ftn@mbersfalse}
\def\paginal{\byp@ge} \def\resetftnumbers{\ftn@mber=1}
\def\ftnote#1{\defm@rk\expandafter\expandafter\expandafter\footnote
    \expandafter\b@@st\m@rk{#1}}

\long\def\jump#1\endjump{}
\def\ssum{\mathop{\lower .1em\hbox{$\textstyle\Sigma$}}\nolimits}

\def\qed{\nobreak\kern 1em \vrule height .5em width .5em depth 0em}
\def\newneq{\hbox{\rlap{\hbox to 1\wd9{\hss$=$\hss}}\raise .1em 
   \hbox to 1\wd9{\hss$\scriptscriptstyle/$\hss}}}
\def\subsetne{\setbox9 = \hbox{$\subset$}\mathrel{\hbox{\rlap
   {\lower .4em \newneq}\raise .13em \hbox{$\subset$}}}}
\def\supsetne{\setbox9 = \hbox{$\subset$}\mathrel{\hbox{\rlap
   {\lower .4em \newneq}\raise .13em \hbox{$\supset$}}}}

\def\vbar{\mathchoice{\vrule height6.3ptdepth-.5ptwidth.8pt\kern-.8pt}
   {\vrule height6.3ptdepth-.5ptwidth.8pt\kern-.8pt}
   {\vrule height4.1ptdepth-.35ptwidth.6pt\kern-.6pt}
   {\vrule height3.1ptdepth-.25ptwidth.5pt\kern-.5pt}}
\def\f@dge{\mathchoice{}{}{\mkern.5mu}{\mkern.8mu}}
\def\b@c#1#2{{\rm \mkern#2mu\vbar\mkern-#2mu#1}}
\def\b@b#1{{\rm I\mkern-3.5mu #1}}
\def\b@a#1#2{{\rm #1\mkern-#2mu\f@dge #1}}
\def\bb#1{{\count4=`#1 \advance\count4by-64 \ifcase\count4\or\b@a A{11.5}\or
   \b@b B\or\b@c C{5}\or\b@b D\or\b@b E\or\b@b F \or\b@c G{5}\or\b@b H\or
   \b@b I\or\b@c J{3}\or\b@b K\or\b@b L \or\b@b M\or\b@b N\or\b@c O{5} \or
   \b@b P\or\b@c Q{5}\or\b@b R\or\b@a S{8}\or\b@a T{10.5}\or\b@c U{5}\or
   \b@a V{12}\or\b@a W{16.5}\or\b@a X{11}\or\b@a Y{11.7}\or\b@a Z{7.5}\fi}}

\catcode`\X=11 \catcode`\@=12
\localtags
\newglobaltag{c}
\def\<#1>{\edef\dotag{\ifmmode\ifinner\message{NO a-TAGS IN EQALIGNNO}
   \else\noexpand\leqno(\bf C\noexpand\ctag{#1})\fi
   \else({\bf C\noexpand\ccite{#1})}\fi}\dotag}
\input amssym.def
\def\bbz{{\Bbb Z}}
\def\bbr{{\Bbb R}}

\def\E#1{\langle#1\rangle}

\def\Ewv#1{\langle w|#1|v\rangle}

\let\vaccent\v
\def\v{|v\rangle}\def\w{\langle w|}

\def\muhat{{\widehat\mu}}\def\mutilde{{\widetilde\mu}}
\def\lambdahat{{\widehat\lambda}}
\def\cite#1{\[#1]}

\catcode`@=11
\def\displaylinesno#1{\displ@y \tabskip\centering
  \halign to\displaywidth{\hfil$\@lign\displaystyle{##}$\hfil\tabskip\centering
    &\llap{$\@lign##$}\tabskip\z@skip\crcr
    #1\crcr}}
\catcode`@=12 
\font \title cmr10 at 14.4pt

 {\nopagenumbers \null
 \vfill 
 \centerline{\title Shift Equivalence of Measures and the Intrinsic Structure of}
 \medskip
 \centerline{\title Shocks in the Asymmetric Simple Exclusion Process}
 \bigskip
 \centerline{B.  Derrida\footnote{${}^1$}{Laboratoire de Physique
Statistique, Ecole Normale Sup\'erieure, 24 rue Lhomond, 75005 Paris, France;
email derrida@lps.ens.fr}, 
S.  Goldstein\footnote{${}^2$}{Department of Mathematics,
Rutgers University, New Brunswick, NJ 08903; email
oldstein@math.rutgers.edu, lebowitz@math.rutgers.edu, speer@math.rutgers.edu},
J.  L.  Lebowitz{${}^{2,}$}\footnote{${}^3$}{Department of Physics,
Rutgers University, New Brunswick, NJ 08903}, and E.  R.  Speer{${}^2$}}

\bigskip \bigskip\bigskip
\centerline{\bf Abstract}
 \medskip
 {\narrower \noindent We investigate properties of non-translation-invariant
measures, describing particle systems on $\bbz$, which are asymptotic to
different translation invariant measures on the left and on the right.  Often 
the structure of the transition region can only be observed
from a point of view which is random---in particular,
configuration dependent.  Two such measures will be called {\it shift 
equivalent} if
they differ only by the choice of such a viewpoint.  We introduce certain
quantities, called {\it translation sums},
which, under some auxiliary conditions, characterize the
equivalence classes.  Our prime example is the asymmetric simple exclusion
process, for which the measures in question describe the microscopic structure
of shocks.  In this case we compute explicitly the translation sums and
find that shocks generated in different ways---in particular, via
initial conditions in an infinite system or by boundary conditions in a
finite system---are described by shift 
equivalent measures.  We show also that when
the shock in the infinite system is observed from the location of a second
class particle, treating this particle either as a first class particle or as
an empty site leads to shift equivalent shock measures.  \par}

\vfill

 \noindent
 {\bf Date:} February 12, 1998
 \medskip\noindent
 {\bf Submitted to:} Issue of {\it Journal of Statistical Physics} dedicated
to Leo Kadanoff
 \medskip\noindent
 \vfill\vfill
\eject}
\pageno1
\openup3\jot

\noindent
 \newsection 1. Introduction

A major component of statistical mechanics, especially its mathematical
aspect, is the study of measures or probability distributions for
infinite particle systems.  Such infinite systems represent
idealizations of macroscopic physical systems whose spatial extension,
although finite, is very large on the microscopic scale of interparticle
distances or interactions.  The advantage of this idealization is that
many phenomena which are clearly manifested in real macroscopic systems,
such as phase transitions, have precise counterparts in the behavior of
the infinite volume measures.  The inevitable boundary and finite size
effects present in real systems, which are frequently irrelevant to the
phenomena of interest, are eliminated in the thermodynamic (infinite
volume) limit \[Ru]. 

Our mathematical characterization of these measures for infinite particle
systems is very good for situations in which the measures are translation
invariant (TI) \[Ru,G].  The situation becomes less transparent when dealing
with spatially nonuniform measures.  These can arise in various ways.  A
rather trivial case occurs when the interaction Hamiltonian or the dynamics
specifying the evolution is position dependent.  More interesting cases arise
when the translation symmetry is broken ``spontaneously'' by the measure.  We
first illustrate these by a well known example from equilibrium statistical
mechanics, then discuss a nonequilibrium example which is the main focus of
this paper. 

Consider the Gibbs measures for the nearest neighbor ferromagnetic Ising
model on ${\Bbb Z}^d$ at low temperatures.  In $d \geq 3$ there exist, in
addition to the TI extremal measures $\mu_+$ and $\mu_-$, in which the
spontaneous magnetizations are $\pm m^*$ with $m^*\ne0$, many non-TI
measures called {\it Dobrushin states}; a family of these can be obtained
as the infinite volume limit of systems with $\pm$ boundary conditions in
the $\hat e_1$-direction \[Dob].  Specifically, let the domain $\Omega$
containing the system consist of sites $j=(j_1,\ldots,j_d)$ 
such that $j_1 \in {\Bbb Z}$ and
$j_2,\ldots,j_d \in [-N,N]^{d-1}$, with all spins outside $\Omega$ being
equal to $+1$ for $j_1 \geq 0$ and $-1$ for $j_1 < 0$.  In each system
configuration we consider the set of $(d-1)$-dimensional surfaces
separating $+1$ and $-1$ spins, formed of $(d-1)$-dimensional faces of
cubes in the dual lattice ${\Bbb Z}^d+(1/2,\ldots,1/2)$.  The {\it
Dobrushin interface} is the maximal connected component of this set which
contains all such $(d-1)$-faces outside $\Omega$.  If $d \geq 3$ then at
low enough temperatures (below the roughening transition) this interface
remains localized near the $j_1=0$ plane as $N$ increases.  Consequently,
in the resulting infinite volume states $\mu^{(\pm)}$ the expectation
values (or correlations) depend on the $\hat e_1$ coordinate, e.g., if
$\sigma_j=\pm1$ is the spin at site $j\in{\Bbb Z}^d$ then
$\langle \sigma_j \rangle_{\mu^{(\pm)}}$ is positive for $j_1 \geq 0$ and
negative for $j_1 < 0$~\[Dob].  This non-translation-invariant infinite volume
Gibbs state is one of an infinite family obtained via translations in the
$\hat e_1$ direction.

If $d=2$, the same boundary conditions produce a translation invariant
state in the infinite volume  limit.  
This is because the interface, while remaining locally sharp,
fluctuates in position with the variance of its displacement from the plane
$j_1=0$ growing like $N$~\[Gal].  
Consequently, the limiting measure (defined by
the $N\to\infty$ limit of local correlation functions) 
is a superposition, with equal weights, of the
extremal translation invariant measures $\mu_+$ and $\mu_-$ \[Gal,MM].
Suppose, however, that we view the system from some point attached to the
Dobrushin interface; for example, we might choose the point $(j^*,0)$,
where $j^*$ is as large as possible so that $(j^*-1/2,0)$ is on the
interface.  Note that the value of $j^*$ and hence the viewpoint will
depend on the configuration under view.  It seems clear that when
$N\to\infty$ a limiting measure will exist which will not be translation
invariant, but will instead approach the state $\mu_+$ (respectively
$\mu_-$) as one goes to infinity in the positive (respectively negative)
$\hat e_1$ direction;  this has not been explicitly
established but for results in this direction see \[BLP].  Other viewpoints
are of course possible, and the resulting measure will depend
in a complicated way on the choice made.
One might choose, for example, $(j^*+j_1,j_2)$ for
some fixed $(j_1,j_2)$, or $(j_*,0)$ with $j_*$ as small as possible so
that $(j_*+1/2,0)$ is on the interface.  One could even add an additional
randomness by choosing either $(j^*,0)$  or $(j_*,0)$ with equal probability;
this seems artificial in the current context but this sort of additional
randomness is natural and necessary in the one-dimensional system to be
studied shortly.

This example illustrates two ways in which
non-TI measures arise.  The non-TI measures for $d\ge3$ arise in viewing
the system from nonrandom, fixed, frames.  Choice of a different frame
simply effects a translation of the measure.  In contrast, the non-TI
measures for $d=2$ can be seen only if one views the system from a
random position---random in the sense that it depends on the
configuration.  Moreover, since the choice of a
viewpoint is rather arbitrary and since the effect on the measure of a
change in viewpoint is hardly transparent, one must now consider a large
family of distinct measures arising from different viewpoints.  Of
course, even in $d\ge3$ we could consider the measure as seen from a
point attached to the interface.  At low temperature there seems to be
little to be gained from such an approach, but it might be of interest
between the roughening and critical temperatures in $d=3$, where the
situation is expected to be similar to that in $d=2$. 

Several questions arise when the transition region must be described by non-TI
measures obtained from configuration-dependent viewpoints.  Is there a
natural choice of viewpoint which will give a best or simplest description of
the local structure of the transition region? How can one extract intrinsic
properties of this region from such a description---properties independent
of the choice of viewpoint? And, given two non-TI measures, how may one
decide if they in fact describe the same system seen from different points
of view? The purpose of the present work is to address such questions.  Our
motivation, and the focus of our study, is in fact not the above example but
a nonequilibrium system, the one dimensional asymmetric simple exclusion
process (ASEP).  The results are, for the moment, also
specific to one-dimensional
systems.

We now describe the non-TI measures arising in the ASEP.  The latter
\[Spi,Ligg] is a model of particles moving on the lattice $\Bbb Z$; a
configuration $\eta$ of the system has the form $\eta =
(\eta(i))_{i\in{\Bbb Z}}$, where $\eta(i)$ is $0$ or $1$ at an empty or
occupied site, respectively.  Dynamically, each particle attempts to
jump to a neighboring site, at random times with rate $1$, choosing its
right or left neighbor with probabilities $p$ and $q$, respectively,
where $p>1/2$ and $q = 1-p$.  The jump takes place if and only if the
target site is empty.  The extremal stationary TI states of this system
are the product (Bernoulli) measures $\nu_\rho$ with constant density
$\rho$ satisfying $0\le\rho\le1$~\[Ligg].  There exist also non-TI
stationary states, which are product measures with nonuniform density
approaching 0 as $i\to-\infty$ and $1$ as $i\to\infty$.  The latter are
in fact special examples of the general class of non-TI measures we will
study here: those which describe the microscopic structure of shocks
present in the macroscopic description of the ASEP. 

The ASEP is described on the (Euler) macroscopic space-time scale by the
inviscid Burgers equation for the particle density $n(x,t) \in
[0,1]$, where $x,t\in\bbr$~(\[R'--'Raz]):
 $${\partial n\over \partial t} 
   + (p-q) {\partial \over \partial x} n(1-n) = 0.\(3) $$
Equation \(3) has shock solutions, $n(x,t) = u(x-Vt)$,  where
$$u(y) = \cases{\rho_- & for $y<0$,\cr
                       \rho_+ &for $y>0$;\cr}\(shsol)$$
here $\rho_+ > \rho_-$ and the velocity is $V = (p-q)(1 - \rho_+ - \rho_-)$. 
A natural question then is what behavior of the ASEP system on the
microscopic level corresponds to this shock solution.  For example, one may
take the initial state $\mu_0$ of the system to be a product measure with
density at site $j$ given by $\rho_-$ for $j<0$ and $\rho_+$ for $j\ge0$, and
ask about the $t\to\infty$ limiting behavior of the state $\mu_t$ at time
$t$.  It might seem that if one were to view the system from a frame moving
with the shock velocity $V$ then one would see in this limit a non-TI state
describing the intrinsic microscopic structure of the shock.  But this is not
true: because fluctuations in the shock position become unbounded 
on the microscopic scale as
$t\to\infty$, the resulting measure is an equal superposition of the product
measures $\nu_{\rho_+}$ and $\nu_{\rho_-}$\[ABL,FF].  It has been shown
\[FKS,F2], however, that there exists a (nonunique) time-dependent random
position $X_t$ such that the $t\to\infty$ limit of the measure $\mu_t$ {\it
seen from the viewpoint $X_t$}, which we shall denote by $\mu'$, exists and
is spatially asymptotic to the product measures $\nu_{\rho_+}$ and
$\nu_{\rho_-}$:
 $$\lim_{k\to\pm\infty}T^{-k}\mu'=\nu_{\rho_\pm}.\(spas)$$
 Here $T$ is the translation operator, which 
acts on configurations by  $(T\eta)(i)=\eta(i-1)$, on functions of
configurations by $(Tf)(\eta)=f(T^{-1}\eta)$, and on measures on
configurations space by $\E{f}_{T\mu}=\E{T^{-1}f}_{\mu}$.  
The situation is thus analogous to that of the two dimensional
Ising model: in the $t\to\infty$ limit here, and in the $N\to\infty$ limit
there, one must look from a configuration-dependent viewpoint to see the
non-TI state.  

The random position $X_t$ discussed above is given by the location of a
single {\it second class particle} inserted into the system, which is then
treated as an empty site in obtaining the measure $\mu'$. This viewpoint is
doubly random, in that the random configuration $\eta$ does not completely
determine the viewpoint, but only its {\it distribution}.  The measure
$\mu'$ is invariant under the ASEP dynamics for the system seen from the
second class particle (we will describe this dynamics below).  In previous
works \[DJLS,DLS] explicit formulas were obtained for a measure $\muhat$
invariant for this same dynamics and with the same spatial asymptotics
\(spas), and it is this measure that will be our main example here;
presumably $\muhat=\mu'$, although this has not been established.  (It is
$\mu'$ which has shown to be obtained by the long time asymptotics
described above.)

In this paper we will focus on questions like those raised above in the
context of the Ising model, which arise from the possibility of different
choices of viewpoint.  In Section~2 we describe the evolution of the ASEP
with a second class particle and the resulting viewpoint on the shock, as
well as several other possible choices of viewpoint.  In Section~3 we
formalize, in a general one-dimensional context, the relation of {\it shift
equivalence} on non-TI measures under which equivalent measures differ by a
random change of viewpoint.  There we define also certain quantities,
called {\it translation sums}, which characterize this equivalence: two
measures (which must satisfy certain additional conditions) are shift
equivalent if and only if all the translation sums for the two measures
agree. This result  will be established in a separate paper \[DGLS2].

In the remainder of the paper we apply these general ideas to the ASEP.  In
Section~4 we utilize the results of \[DLS] to compute the translation sums
explicitly.  From this computation (and using the verification, here omitted,
that the ASEP shock measures satisfy the additional conditions mentioned
above) we establish both negative and positive results about the ASEP shock. 
In Section~5 we show that for certain values of the parameters the shock
measure is not shift equivalent to any product measure with a monotone
density; we show also that when the shock is observed from the location of a
second class particle, treating this particle either as a first class
particle or as an empty site leads to shift equivalent shock measures. 
Finally, in Section~6 we show that certain shocks arising in versions of the
ASEP with different boundary conditions are in fact shift equivalent. 

 \newsection 2. Points of view for the ASEP shock
 
In this section we illustrate the nature of shock measures by considering
various viewpoints on the shock for the ASEP, beginning with the viewpoint
from a second class particle.  We first describe briefly the properties of
the ASEP when a single second class particle is introduced into the system
\[ABL].  The second class particle has its own dynamics: it attempts to
jump exactly as does an ordinary (first class) particle, succeeding only if
the target site is empty; on the other hand, when a first class particle
attempts to jump onto the site occupied by the second class particle, the
jump succeeds and the two particles exchange sites.  A configuration of
this system is $\tau = (\tau(i))_{i\in{\Bbb Z}}$, where $\tau(i)$ is $0$ if
site $i$ is unoccupied, $1$ if it is occupied by one of the original
particles, now called {\it first class particles}, and $2$ if it is
occupied by the second class particle.  Let us denote the location of the
second class particle by $X$.  If $\lambda_t$ is a measure on this system
evolving under the above dynamics and $X_t$ the corresponding location of
the second class particle at time $t$, we write $T^{-X_t}\lambda_t$ for the
measure describing the configurations as  seen from the second class particle.

There is an alternate, equivalent way to describe the system with a second
class particle \[ABL].  Consider two copies of the ASEP system having
configurations $\eta_0$ and $\eta_1$ which agree except at one site $X$, at
which $\eta_0(X)=0$ and $\eta_1(X)=1$, i.e., system~0 has a hole and system~1
a particle.  Allow this pair of systems to evolve under a coupled dynamics,
so that attempts to jump from a given site to an adjacent one occur
simultaneously.  Then each time a jump occurs in either system the same jump
occurs in the other, if possible; this synchronization can fail only when the
extra particle in system~1 jumps, or when a particle in system~0 jumps on the
extra hole, and in these cases the mismatch position $X$ will move.  From a
configuration $(\eta_0,\eta_1)$ of this doubled system we may obtain a
configuration $\tau$ of the single ASEP with second class particle by taking
$\tau(X)=2$ and $\tau(i)=\eta_0(i)=\eta_1(i)$ when $i\ne X$; the dynamics for
the doubled ASEP system corresponds to that described above of the system
with a second class particle.  Conversely, from a configuration $\tau$ we may
obtain two distinct ASEP configurations $\eta_0$ and $\eta_1$ by restricting
attention to one or the other of the paired ASEP systems or equivalently by
replacing the second class particle by respectively a hole or a first class
particle; we will write $\eta_0=\Psi_0(\tau)$ and $\eta_1=\Psi_1(\tau)$. 
Similarly, from a measure $\lambda$ for the ASEP with second class particle
we obtain ASEP measures $\Psi_0(\lambda)$ and $\Psi_1(\lambda)$ giving the
distribution of $\eta_0$ and $\eta_1$ under $\lambda$.  Clearly if
$\lambda_t$ is evolving under the second class particle dynamics then both
$\Psi_0(\lambda_t)$ and $\Psi_1(\lambda_t)$ evolve under the simple ASEP
dynamics. 

A measure $\lambdahat$ describing this system from the viewpoint of the
second class particle, i.e., in a reference frame in which $X=0$, was
constructed explicitly in \[DLS]; the construction is summarized in
Section~4.  This measure has spatial asymptotics corresponding to the shock,
 $$\lim_{k\to\pm\infty}T^{-k}\lambdahat=\nu_{\rho_\pm},\(lamas)$$
and is invariant under the natural dynamics for the system seen from the
second class particle, under which the second class particle is always at the
origin and a jump of this particle in the original dynamics becomes a jump of
the rest of the system in the opposite direction.  

To obtain a measure on the original ASEP configurations from the measure
$\lambdahat$ we may consider either $\Psi_0(\lambdahat)$ or
$\Psi_1(\lambdahat)$.  To be definite, let us focus for the moment on the
former, which gives the distributions of the configuration $\eta_0$, and
denote it by $\muhat$; $\muhat$ is obtained from $\lambdahat$ by replacing
the second class particle at the origin with a hole.  This is the measure
referred to in the introduction.  We may allow it to evolve to $\muhat_t$
under the ASEP dynamics, or equivalently write
$\muhat_t=\Psi_0(\lambdahat_t)$; then viewed from the position $X_t$ it is
time invariant:
 $$T^{-X_t}\muhat_t=\muhat.\(linvmu)$$
 Thus $\muhat$ furnishes an invariant description of the shock itself,
ignoring its location.  It follows from \(lamas) that $\muhat$ has the
spatial asymptotics \(spas).   

Other points of view are possible.  Thus one may obtain new descriptions of
the shock by rather trivial shifts of viewpoint; for example, a constant
one, to the second site to the right of the second class particle, a
configuration-dependent shift, to the third empty site to its right, or a
shift with additional randomness, such as a choice, with equal weights,
between the two previous possibilities.

An alternate measure for the shock is implicit in the description of the
system with second class particle as a coupled pair of ASEP systems: the
measure $\mutilde_t=\Psi_1(\lambdahat_t)$, obtained from $\lambdahat_t$ by
replacing the second class particle by a first class particle, evolves with
the ASEP dynamics, and is invariant in the sense of \(linvmu):
$T^{-X_t}\mutilde_t=\mutilde_0$.  We write
$\mutilde=\mutilde_0=\Psi_1(\lambdahat)$; $\mutilde$ is obtained from
$\muhat$ simply by replacing the empty site at the origin by a particle.
Despite the fact that $\mutilde$ and $\muhat$ are equally valid as
candidates for the description of the ASEP shock (in the terminology
introduced in the next section, they are both invariant shock measures for
the ASEP), it is not at all clear that they are shift equivalent, that is,
differ by the sort of random change of viewpoint that we have been
considering here.  In Section~5 we will use the ideas of the next section
to show that this is the case.

Other choices may for some purposes be more tractable.  The existence of a
shock measure was first proved by Ferrari, Kipnis, and Saada \[FKS] (FKS)
using a random viewpoint $Z_t$ which Ferrari \[F2] later showed was related
to $X_t$ by a random translation of finite mean.  To construct $Z_t$,
consider again two copies of the ASEP system with configurations $\zeta_0$
and $\zeta_1$ satisfying $\zeta_0(k)\le\zeta_1(k)$ for all $k$, so that
when there is a particle at site $k$ in configuration $\zeta_0$ there is
also a particle at that site in configuration $\zeta_1$.  Allow the system
to evolve under the coupled dynamics described above, so that again sites
$k$ at which $\zeta_0(k)=\zeta_1(k)=1$ and those at which $\zeta_0(k)=0$
and $\zeta_1(k)=1$ obey the dynamics of first and second class particles,
respectively.  Let $\lambda^*$ be a translation- and time-invariant measure
for this system in which the densities are given by
$\E{\zeta_0(0)}_{\lambda^*}=\rho_-$ and $\E{\zeta_1(0)}_{\lambda^*}=\rho_+$
(the existence of such a $\lambda^*$ is established in \[FKS]).  Randomly
select some second class particle, i.e., some discrepancy between $\zeta_0$
and $\zeta_1$ (more precisely, condition on the presence of
such a particle at the origin at time 0), and let $Z_t$ be its position at
time $t$.  At time $t=0$, define an ASEP configuration $\eta$ as follows:
first, if $\zeta_0(k)=\zeta_1(k)=1$ then $\eta(k)=1$, and if
$\zeta_0(k)=\zeta_1(k)=0$ then $\eta(k)=0$; second, if $\zeta_0(k)=0$ and
$\zeta_1(k)=1$, and $k$ is the $j^{\rm th}$ site at which such a
discrepancy occurs, counting from $j=0$ at $Z_0$, then $\eta(k)=1$ with
probability $1/(1+(q/p)^j)$, and $\eta(k)=0$ with the complementary
probability, and all of these choices are independent.  Allow the
configuration $\eta$ to evolve with ASEP dynamics coupled to that of
$\zeta_0$ and $\zeta_1$, so that if a jump occurs in any of the three
systems it also occurs in any others in which it is possible.  Then the
distribution of $\eta$ at time $t$, viewed from the position $Z_t$, is time
independent and has the shock asymptotics \(spas).

In all the examples considered so far the position of the viewpoint is
not determined by knowledge of the ASEP configuration $\eta$: knowing
$\eta$ gives only the {\it distribution} of the random viewpoint.  This is an
additional randomness beyond the configuration dependence discussed for
the $d=2$ Ising model in the introduction.  Our last example is a
construction of a viewpoint $\ell$ on a non-TI measure $\mu$ which, as
in the $d=2$ equilibrium example, is intrinsic.  This means, first, that
the viewpoint $\ell(\eta)$ depends only on the configuration $\eta$,
with no additional randomness, and second, that the viewpoint behaves
covariantly under translations, so that 
 $$\ell(T\eta)=\ell(\eta)+1.\(covar)$$
   The function $\ell(\eta)$ may be thought of as picking out a shock
location in the configuration $\eta$.  Its intrinsic nature means that if
$\nu$ is any measure obtained from $\mu$ by a shift of viewpoint then $\nu$
and $\mu$ look the same from the configuration dependent viewpoint $\ell$.
In the construction of $\ell$, $\mu$ need not be related to the ASEP
dynamics; we require only that $\mu$ be a non-TI measure on ASEP
configurations which has well-defined and configuration-independent
asymptotic densities satisfying $\rho_+>\rho_-$:
 $$\rho_\pm=\lim_{N\to\pm\infty}{1\over |N|+1}\sum_{k=0}^N \eta(k),\(asden)$$
 for $\mu$-almost every $\eta$.
 
To define $\ell(\eta)$ we first define a function $h_\eta(j)$ to be the
signed cumulative occupation from the origin to site $j$:
 $$h_\eta(j)=\cases{\sum_{i=1}^j\eta(i),& if $j>0$,\cr\noalign{\vskip5pt}
   0,& if $j=0$,\cr\noalign{\vskip5pt}
  -\sum_{i=j+1}^0\eta(i),& if $j<0$.\cr}\(defh)$$
 The graph of $h_\eta$ has, for typical $\eta$, slope $\rho_+$ (on a large
scale) far to the right of the origin and $\rho_-$ far to the left.  Now fix
an irrational number $\rho_*$ satisfying $\rho_-<\rho_*<\rho_+$, and for any
$\eta$ define $\ell(\eta)$ to be the integer $j$ which minimizes
$h_\eta(j)-\rho_*j$, whenever such a (necessarily unique) minimizing integer
exists; for other $\eta$, $\ell(\eta)$ is undefined.  The special role played
by the origin in the definition \(defh) of $h_\eta$ does not affect the value
of $\ell(\eta)$.  The construction is shown graphically in Figure~1.  It is
intuitively clear, and can be proved, that $\ell$ is well defined and finite
with probability one relative to $\mu$.  The definition of $\ell$ depends
strongly on the choice of $\rho_*$; slight changes in $\rho_*$ will cause
large changes in the viewpoint $\ell(\eta)$ for some configurations $\eta$. 

Note that the sets $S_k=\{\,\eta\mid\ell(\eta)=k\,\}$ form a partition of the
configuration space (up to a set of $\mu$-measure zero) which is nicely
mapped by translations: $T(S_k)=S_{k+1}$.  Such partitions have been
constructed for more general $T$ in the context of discrete time dynamical
systems by Gurevi\vaccent c and Oseledec \[O]. 

The shock may look quite different from different viewpoints, in particular,
from viewpoints $\ell$ defined with different values of $\rho_*$.  In
Figure~2 we show shock profiles (mean values $\E{\eta(k)}$ at site $k$
relative to the viewpoint adopted) for the shock in the totally asymmetric
($p=1$) model with densities $\rho_+=0.7$, $\rho_-=0.2$, seen from three
different viewpoints: from the second class particle (that is, in the measure
$\muhat$) and from $\ell_1$ and $\ell_2$, the viewpoints constructed as above
with $\rho_{*1}=\pi^{-1}\rho_+ + (1-\pi^{-1})\rho_-$ and
$\rho_{*2}=(1-\pi^{-1})\rho_+ + \pi^{-1}\rho_-$. 

 \newsection 3. Equivalence of measures under random shifts

  In the previous section we have described implicitly an equivalence
relation on probability measures on the ASEP configuration space
$S=\{0,1\}^{\bbz}$, under which two such measures $\mu_1$ and $\mu_2$ are
equivalent if they differ by a configuration dependent random shift of
viewpoint.  Perhaps the simplest way to make this precise is in terms of a
{\it coupling} for the two measures, that is, a measure on $S\times S$ with
marginals $\mu_1$ and $\mu_2$ on the first and second components.  We say
that $\mu_1$ and $\mu_2$ are {\it shift equivalent} if there exists such a
coupling $\mu^*$ and an integer-valued function $Y$ such that for
$(\eta_1,\eta_2)\in S\times S$, $\eta_2=T^{-Y(\eta_1,\eta_2)}\eta_1$ with
$\mu^*$-probability one; in this case we write
 $$\mu_2=T^{-Y}\mu_1. \(equiv)$$
 Generalizations to measures on sets other than $S$ on which a translation
operator acts can easily be made.  The coupling $\mu^*$ is sometimes
referred to in the mathematical literature as a {\it shift coupling}
\[ATh,Th] or an {\it orbit coupling} \[G2]. 

  It is sometimes convenient to work with alternate formulations of this
equivalence relation.  One such is based on a variation of the well known
Vasershtein distance \[Vass] between two measures.  Define $d(\eta_1,\eta_2)$
on $S\times S$ to be the minimum, over $k$, of the number of sites at which
$\eta_1$ and $T^k\eta_2$ differ, and for measures $\mu_1$ and $\mu_2$ on $S$
define $D(\mu_1,\mu_2)$ by
 $$D(\mu_1,\mu_2)=\inf_{\nu^*}\E{d}_{\nu^*},\(defD)$$
 where the infimum is over couplings $\nu^*$ for $\mu_1$ and $\mu_2$.  Then
it can be shown \[DGLS2] that $\mu_1$ and $\mu_2$ are shift equivalent 
if and only if $D(\mu_1,\mu_2)=0$.  The minimum in \(defD) is in
fact achieved when $\nu^*=\mu^*$, with $\mu^*$ the coupling used to define
\(equiv). 

A second reformulation of the relation of shift equivalence 
is obtained by noting 
that if \(equiv) holds then clearly
 $$\mu_1(A)=\mu_2(A) \qquad
      \hbox{for all $A\subset S$ and $T(A)=A$}, \(equiv2)$$
 that is, if $A$ is translation invariant.  Conversely, it can be shown
\[Th,G2] that if \(equiv2) holds then there exists a random position $Y$ so
that \(equiv) is satisfied.  For measures describing shocks,
``interesting'' TI sets $A$ with nontrivial probability describe intrinsic
properties of the shock; for example, we might take $A$ to be the set of
configurations $\eta$ with a particle at the site three sites ahead of the
position $\ell(\eta)$ defined in Section~2, so that
$\mu(A)=\E{\eta(\ell(\eta)+3)}_\mu$.  

 Finally, as remarked earlier, when an intrinsic viewpoint $\ell(\eta)$ can
be defined, that is, a function $\ell(\eta)$ defined almost everywhere with
respect to both $\mu_1$ and $\mu_2$ and satisfying \(covar), then $\mu_1$ and
$\mu_2$ are shift equivalent if and only if $T^{-\ell}\mu_1=T^{-\ell}\mu_2$. 

 The concept of shift equivalence allows us to give precise definitions of
two natural concepts for shock measures in the ASEP (or similar systems).  We
say that a measure $\mu$ is an {\it invariant shock measure} if it has the
spatial asymptotics \(spas) for some $\rho_\pm$ and if, when $\mu_t$ is the
measure evolving under the ASEP dynamics which satisfies $\mu_0=\mu$, $\mu_t$
is shift equivalent to $\mu$ for all $t$;  for example, the measures $\muhat$
and $\mutilde$ of the previous section 
are invariant shock measures in this sense.  We say that the ASEP has a {\it
unique} shock measure (for given $\rho_\pm$) if any two such invariant
measures are shift equivalent.  It seems natural to conjecture that the ASEP
has a unique shock measure in this sense for all $\rho_\pm$ satisfying
$\rho_+>\rho_-$, but this has not been established. 

Let us now restrict attention to measures on $S$ which, like the ASEP shock
measures considered in Section~2, converge under spatial translation to
distinct TI states.  Suppose then that $\mu_\pm$ are translation invariant
probability measures on $S$ with $\mu_+\ne\mu_-$, and define
a {\it ramp measures} to be a 
probability measure $\mu$ on $S$ which is asymptotic to $\mu_+$ to the
right of the origin and to $\mu_-$ to the left:
 $$\lim_{k\to\pm\infty}T^{-k}\mu=\mu_\pm.\(asymp)$$

Associated to each ramp measures is a family of {\it translation sums}. 
These sums, under rather mild additional technical conditions on the measures
involved \[DGLS2], are invariant under a shift of viewpoint and furnish a
complete characterization of shift equivalence.  In a sense, the
equivalence of \(equiv) and \(equiv2) also provides a set of invariant
quantities which determine the equivalence class of a measure $\mu$: the
values $\mu(A)$ for all TI sets $A$.  The example of the translation
invariant set given above, however, suggests correctly that these quantities
are difficult to calculate.  We will see in the next section that the
translation sums are calculable for the ASEP shock measure. 

Suppose that $\mu$ is a ramp measure and that $f$ is a function on $S$ which
depends on only finitely many occupation numbers and which satisfies
$\mu_+(f)=\mu_-(f)=0$; for example, $f(\eta)=\eta(1)-\eta(0)$ is such a
function and, if the asymptotic states are product measures, i.e., if
$\mu_{\pm}=\nu_{\rho_\pm}$, then so is $f(\eta)=\eta(k)(\eta(1)-\eta(0))$
whenever $k\ne0,1$.  Then we may define the {\it translation sum}
 $$\Delta_\mu(f)=\sum_{n=-\infty}^\infty\langle T^nf\rangle_{\mu};\(bsum)$$
 $\Delta_\mu(f)$ will be finite for any ramp measure $\mu$ for which the
asymptotic behavior \(asymp) is achieved with sufficient rapidity to
guarantee that this sum converges.  For example, if $f(\eta)=\eta(1)-\eta(0)$
then $\Delta_\mu(f)=\rho_+-\rho_-$ since \(bsum) telescopes, and similarly
$\Delta_\mu(f)=\E{g}_{\mu_+}-\E{g}_{\mu_-}$ if $f=Tg-g$ for some $g(\eta)$. 
However, we see no easy way to compute $\Delta_\mu(f)$ for
general $\mu$ and $f$.  

The values of the translation sums characterize the (shift) equivalence
classes of ramp measures in the following sense: under additional
conditions describing the convergence at $\pm\infty$, two ramp measures
$\mu_1$ and $\mu_2$ are shift equivalent if and only if
 $$\Delta_{\mu_1}(f)=\Delta_{\mu_2}(f)\(equivbd)$$
 for all $f$ satisfying $\mu_+(f)=\mu_-(f)=0$ \[DGLS2]. 

As a simple application of this result, consider a product measure $\nu_1$ on
$S$ with density $\rho_1(k)=\E{\eta(k)}_{\nu_1}$ satisfying
$\lim_{k\to\pm\infty}\rho_1(k)=\rho_\pm$, with $\rho_+\ne\rho_-$; $\nu_1$ is
a ramp measure (asymptotic to $\nu_{\rho_+}$ and $\nu_{\rho_-}$) and the
additional technical considerations needed for the above result are satisfied
if the asymptotic limit is achieved sufficiently rapidly.  Let $\nu_2$ be
another such measure obtained by altering the density at the origin only:
$\rho_2(0)\ne\rho_1(0)$, $\rho_2(k)=\rho_1(k)$ if $k\ne0$.  Then $\nu_1$ and
$\nu_2$ will typically have different translation sums and hence not be shift
equivalent; for example, consideration of the translations sums for the
functions $f_k(\eta)=\eta(k)(\eta(1)-\eta(0))$, $k>1$, shows that $\nu_1$ and
$\nu_2$ will not be shift equivalent unless $\rho_1(k)+\rho_1(-k)$ is
independent of $k$.

 \newsection 4. Translation sums for the ASEP shock measure

  In this section we show how to compute the translation sums 
$\Delta_\muhat(f)$
for the ASEP shock measure $\muhat$ which, as discussed in Section 2, is
obtained from the invariant shock measure $\lambdahat$ for the system with
second class particle at the origin by replacing that particle by a hole.  In
this case the asymptotic measures $\mu_{\pm}$ are the product measures
$\nu_{\rho_\pm}$.  We will continue to denote a typical ASEP configuration by
$\eta$ ($\eta(i)=0$ or $1$) and a configuration in which second class
particles may occur by $\tau$ ($\tau(i)=0$, $1$, or $2$); in this section
such a $\tau$ will always contain a single second class particle located at
the origin. 

In \cite{DLS} (which was an extension to the general ASEP of the results of
\cite{DJLS} for the totally asymmetric model, in which $p=1$) it was shown
that the measure $\lambdahat$ can be written in terms of two vectors $\v$ and
$\w$ and three operators $A$, $D$ and $E$ satisfying the following algebraic
rules:
 $$\eqalignno{
 p D E -q E D
  &= (p-q) [(1- \rho_-)(1-\rho_+) D + \rho_- \rho_+ E ],&\(alg1)\cr
 p A E  -q E A 
  &= (p-q)(1-\rho_-)(1-\rho_+) A ,&\(alg2)\cr
 p D A-q A D 
  &= (p-q) \rho_+ \rho_- A ,& \(alg3)\cr
  (D + E ) | v \rangle &= |v \rangle, &\(alg4)\cr
  \langle w | (D + E ) &= \langle w | , &\(alg5)\cr
 \langle w | A | v \rangle &= 1 . &\(alg6)\cr}
 $$
 Specifically, the probability of the set of configurations specified by the
occupation numbers $\zeta(i)$ ($\zeta(i)=0,1$) of $m$ consecutive sites to
the left of the second class particle (which is located at the origin) and
$n$ consecutive sites to its right,
 $$\lambdahat\bigl(\{\,\tau\mid
       \tau(i)=\zeta(i),\ i=-m,\ldots,-1\hbox{ and }i=1,\ldots,n\,\}\bigr)
  \equiv P_{m,n}(\zeta),\(forjoel)$$
 can be written as the matrix element in
which a first class particle is represented by a matrix $D$, an empty site
by a matrix $E$, and the second class particle by a matrix $A$:
 $$
 P_{m,n}(\zeta)=
   \Ewv{\left\{\prod_{i=-m}^{-1} [ \zeta(i) D+ (1 -\zeta(i)) E] \right\}
  A \left\{\prod_{j=1}^n[\zeta(j) D + (1 -\zeta(j)) E ] \right\}}. \(weight)
 $$
 For example, the probability of finding occupation numbers $1\;0\;1$
immediately to the left and $0\;1\;1\;0\;0$ immediately to the right of
the second class particle, that is, of the local configuration
$1\;0\;1\;2\;0\;1\;1\;0\;0$, is given by $\Ewv{DEDAED^2E^2}$. 

It follows from the above that the microscopic shock profile, defined as the
average occupation $\E{\tau(n)}_\lambdahat=\E{\eta(n)}_\muhat$ at
site $n\ne0$, is given for $n>0$ by
 $$\langle\tau(n)\rangle_\lambdahat = \Ewv{A(D + E)^{n-1}D}.\(taun)$$
 The exact expression of this profile was given in equations (4.1--4.5) of
\cite{DLS}, where it was also shown that the profile has the symmetry
property
 $$\E{\tau(n)}_\lambdahat+\E{\tau(-n)}_\lambdahat=\rho_++\rho_-.\(sym)$$

 Let us call a finite product of $D$'s and $E$'s a {\it word}.  To every
function of $k$ consecutive site occupation numbers, say
$f(\eta(1),\eta(2),...,\eta(k))$, there is naturally associated a linear
combination of words of length $k$; for example, if
 $f= \eta(1)(1- \eta(3)) + 3\eta(5)\eta(6)$ then
 $$W=D (D+E)E (D+E)^3 + 3 (D+E)^4 D^2. \(corr)$$
 We denote the word of length zero by $1$ and associate it with the
function $f=1$.  For every linear combination $W$ of words we will define
below a number $\Gamma(W)$ and will then develop, from the algebraic rules
\(alg1)--\(alg6), new rules which enable us to calculate $\Gamma(W)$.  This
will determine the translation sums
$\Delta_{\muhat}(f)$ defined in \(bsum), since we will show below that when
$f$ and $W$ are related as above and in addition $\E{f}_{\mu_\pm}=0$,
$\Delta_\muhat(f)=\Gamma(W)$. 

 Suppose then that $W$ is a linear combination of words and define
 $$ \Gamma(W)= \lim_{L,M\to\infty}
  \left\{\Gamma_{L,M}(W)-L\,r_+(W)-M\,r_-(W) \right\}. \(CW) $$
 Here
 $$ \Gamma_{L,M}(W) = \left.  {d \over d \theta}
   \Ewv{(\tilde{D}+\tilde{E})^L \tilde{W} (\tilde{D}+\tilde{E} )^M}
   \right|_{\theta=0}, \(CLMW) $$
 with
 $$ \tilde{D}= D, \qquad\qquad \tilde{E} = E  + \theta A ,
 \(tilde) $$
 and $\tilde{W}$ the operator obtained from $W$ by replacing each $D$
and $E$ in $W$ by $\tilde{D}$ and $\tilde{E}$; $r_\pm(W)$  denotes
the number obtained by replacing each $D$ and $E$ in $W$ by $\rho_\pm$
and $1-\rho_\pm$ respectively.  It is clear from \(CLMW) and the
definition \(tilde) of $\tilde{D}$ and $\tilde{E}$ that the right hand
side of \(CLMW) is a sum of matrix elements of products of the operators
$D$, $E$, and $A$, with each product containing a single operator $A$;
each such matrix element is computable from \(alg1)--\(alg6) and
represents a probability calculated in the measure $\lambdahat$.  Moreover,
the limit in \(CW) exists because $\lambdahat$ converges
exponentially fast to $\nu_{\rho_\pm}$ at $\pm \infty$ \[DLS].  Thus
$\Gamma(W)$ is well defined. 

It follows directly from the definition of $\Gamma(W)$ that
 $$\eqalignno{\Gamma(a_1 W_1 + a_2 W_2) &= a_1 \Gamma(W_1) + a_2 \Gamma(W_2),
  \qquad \hbox{for any   $a_1$ and $a_2$;}&\(ALG1)\cr
 \Gamma( W [D+E]) &= \Gamma(W) +  r_-(W); & \(ALG2) \cr
 \Gamma(  [D+E] W) &= \Gamma(W) +  r_+(W);& \(ALG3)\cr
 \Gamma( 1) &= 0. &\(ALG4) \cr} $$
Also, one can easily show that $\tilde{D}$ and
$\tilde{E}$ satisfy \(alg1), and hence for any $W_1$ and $W_2$,
 $$ \Gamma(W_1(p D E -q E D)W_2)
 = (p-q) \Gamma(W_1[(1- \rho_-)(1-\rho_+) D + \rho_- \rho_+ E]W_2).\(ALG5) $$

 The five rules \(ALG1)--\(ALG5), together with the value of $\Gamma(D)$,
 which we will derive below, allow one to calculate $\Gamma(W)$ for any
linear combination of words $W$.  The argument, based on a recursion on the
length of a word $W$, is almost identical to that given in \cite{Sandow,DLS}
to show that \(alg1)--\(alg6) are sufficient to calculate any matrix element
of the form \(weight).  Consider, for example, $\Gamma(DW_n)$, where $W_n$ is
a word of length $n\ge1$ with $k$ factors $E$:
 $$\eqalignno{
  p^n\Gamma(DW_n)
  &= q^kp^{n-k}\Gamma(W_nD) + \hbox{l.o.t.}\cr
  &= -q^kp^{n-k}\Gamma(W_nE)+q^kp^{n-k}r_-(W_n) + \hbox{l.o.t.}\cr
  &= -q^n\Gamma(EW_n) +q^kp^{n-k}r_-(W_n) + \hbox{l.o.t.}\cr
  &= q^n\Gamma(DW_n) - q^nr_+(W_n) + q^kp^{n-k}r_-(W_n) + 
          \hbox{l.o.t.}&\(reduce)\cr}
 $$
 Here l.o.t.  (lower order terms) denotes linear combinations of $\Gamma(W)$
for words $W$ of length $n$ or less.  Since $q<p$, \(reduce) can be solved
for $\Gamma(DW_n)$.  For example, since \(ALG2) and  \(ALG4) imply
that $\Gamma(D+E)=1$,
 $$\eqalignno{p\Gamma(D^2)&=-p\Gamma(DE)+p\Gamma(D)+p\rho_-\cr
  &= -q\Gamma(ED)-(p-q)\Gamma((1-\rho_+)(1-\rho_-)D+\rho_-\rho_+E) 
        +p\Gamma(D)+p\rho_-\cr
  &= q\Gamma(D^2) + p \rho_- - q \rho_+ -(p-q)\rho_-\rho_+
      + (p-q)(\rho_++\rho_-)\Gamma(D),&\(GamD2)\cr}$$
 and hence 
 $$
 \Gamma(D^2)=  {p \rho_- - q \rho_+ \over p -q} - \rho_+ \rho_- 
    + (\rho_++\rho_-)\Gamma(D).\(D2) $$

 If $W$ is a word of length $k$ which contains $j$ factors $E$ and if for
$i=1,\ldots,j$, $W^{(i)}$ is the operator product obtained from $W$ by
replacing the $i^{\rm th}$ factor $E$ by $A$, then from \(alg4), \(alg5),
\(CLMW), and \(tilde),
 $$\Gamma_{L,M}(W)
   =\sum_{i=0}^{L-1}\Ewv{A(D+E)^iW}  +\sum_{i=1}^j\Ewv{W^{(i)}} 
         +\sum_{i=0}^{M-1}\Ewv{W(D+E)^iA}.  \(gam)$$
 More generally, if $f(\eta(1),\eta(2),... ,\eta(k))$ is a function of the
occupation numbers of sites $1,\ldots,k$ and $W$ the corresponding linear
combination of words of length $k$ (see \(corr)), then \(gam) implies that
 $$\Gamma_{L,M}(W)=\sum_{i=-M-k}^{L-1}\E{T^if}_{\hat\mu}\;;\(fun)$$
 Note that, as is clear in \(gam), the averages are taken in the measure
$\muhat$ obtained from $\lambdahat$ by replacing the second class particle at
the origin by a hole, that is, taking $\eta(0)=0$.  For example, if $f=
\eta(1)(1- \eta(3)) + 3\eta(5)\eta(6)$ then from \(corr) and \(fun),
 $$\Gamma_{L,M}(D (D+E)E (D+E)^3 + 3 (D+E)^4 D^2) = \sum_{ i=-M-5}^{L} 
  \E{\eta(i)(1-\eta(i+2)) + 3\eta(i+4)\eta(i+5)}_\muhat.
   \(gamex) $$

 Equations \(fun) and \(sym) imply that
 $$\Gamma(D)=\lim_{L\to\infty}
  \left\{ \sum_{i=-L}^L\E{\eta(i)}_{\muhat}-L(\rho_++\rho_-)\right\}
  = \E{\eta(0)}_\muhat=0.\(ALG6bis)$$
 Moreover, \(fun) leads immediately to the calculation of translation sums
$\Delta_\muhat(f)$, for if $f$ is a function satisfying $\E{f}_{\mu_\pm}=0$
and $W$ is the corresponding linear combination of words, then
\(CW), \(fun), and the relations $\E{f}_{\nu_{\rho_\pm}}=r_\pm(W)$ imply that
 $$\Delta_\muhat(f)=\Gamma(W). \(Del)$$

The translation sums
are in some cases related to expectation values in the measure
$\muhat$ in a surprisingly simple way.  For example, let $f_1(\eta)$ be a
function of $\eta(-m),\ldots,\eta(-1)$ and $f_2(\eta)$ a function of
$\eta(2),\ldots,\eta(n)$, and let $h(\eta)=\eta(1)-\eta(0)$.  Then
$\mu_\pm(f_1hf_2)=0$, so that $\Delta_\muhat(f_1hf_2)$ is defined.  We will
show that
 $$\Delta_\muhat(f_1hf_2)
=(\rho_+-\rho_-)\E{f_1(T^{-1}f_2)}_\muhat.\(surprise)$$

 The operator corresponding to $h$ is $A'=(D+E)D-D(D+E)=ED-DE$; in view of
\(weight) and \(Del), \(surprise) will
follow if we show that for any $W_1$, $W_2$,
 $$\Gamma(W_1A'W_2) =(\rho_+-\rho_-)\Ewv{W_1AW_2}.\(sur2)$$
 To verify \(sur2) we show that, up to a normalization, $\Gamma(W_1A'W_2)$
is determined by the same algebraic rules \(alg1)-\(alg6) which determined
$\Ewv{W_1AW_2}$.  First, $D$ and $E$ of course satisfy \(alg1), and this in
turn implies that $A'$ satisfies the analogue
of \(alg2) and \(alg3):
 $$\eqalignno{
 p A' E  -q E A' 
  &= (p-q)(1-\rho_-)(1-\rho_+) A' ,&\(alg2')\cr
 p D A'-q A' D 
  &= (p-q) \rho_+ \rho_- A' .& \(alg3')\cr}$$
 Moreover, since $r_+(W_1 A' W_2)= r_-(W_1 A' W_2)=0$ for any $W_1,W_2$,
\(ALG2) and \(ALG3) give
$$\Gamma([D+E] W_1 A' W_2) = \Gamma( W_1 A' W_2 [D+E])
     = \Gamma( W_1 A'W_2),\(alg45') $$ 
 which is the analogue of \(alg4)--\(alg5).  Finally, the analogue of
\(alg6) is
 $$\Gamma(A') = \Gamma([D+E]D)-\Gamma(D[D+E]) =\rho_+-\rho_-.\(alg6')$$
Since the rules \(alg1)--\(alg6) determine all matrix elements
$\Ewv{W_1AW_2}$, \(alg1) and \(alg2')--\(alg6') imply that these will agree
with the $\Gamma(W_1A'W_2)$ up to a normalization determined by comparing
\(alg6) and \(alg6'), that is, \(sur2) holds.

 \newsection 5. Consequences of the calculation of the translation sums

 In this section we apply the calculation of the translation sums
$\Delta_\muhat(f)$ outlined in Section~4 to discuss the
shift equivalence of various measures describing the ASEP
shock and to identify certain intrinsic features of the ASEP shock.

 In Section~2 we observed that the measure $\mutilde$, obtained from
$\lambdahat$ by replacing the second class particle at the origin by a
first class particle (or equivalently from $\muhat$ by replacing the
hole at the origin by a particle), stood on an equal footing with
$\muhat$ in providing a description of the shock.  We now show that
 $$\Delta_\muhat(f)=\Delta_\mutilde(f)\(same)$$
 for any $f$ satisfying $\E{f}_{\mu_\pm}=0$, so that by the general results
referred to in Section~3, $\muhat$ and $\mutilde$ differ only by a shift of
viewpoint (which in general will be a random shift with distribution
depending on the configuration).

For suppose that, instead of using \(tilde), we had defined $\tilde{D}$ and
$\tilde{E}$ by $\tilde{D}= D +\theta A $ and $\tilde{E} = E $, arriving at
new quantities $\tilde\Gamma(W)$.  Then all the considerations of Section~4
would have been essentially unchanged, except that averages in \(fun) would
be computed in the measure $\mutilde$; as a result, \(ALG6bis) would become
$\tilde\Gamma(D)=1$.  This would lead to
 $$\eqalignno{\tilde\Gamma(W) 
  &= \Gamma(W)
     + [\tilde\Gamma(W) - \Gamma(W)]{r_+(W) - r_-(W) \over \rho_+ - \rho_-}\cr
  &= \Gamma(W)+ {r_+(W) - r_-(W) \over \rho_+ - \rho_-};&\(newGam)}$$ 
 this equation may be verified by noting that
$\Gamma_1(W)\equiv \tilde\Gamma(W) -\Gamma(W)$ and
$\Gamma_2(W)\equiv r_+(W)-r_-(W)$ both satisfy \(ALG1), \(ALG4), \(ALG5),
and the homogeneous versions of \(ALG2) and \(ALG3), and that \(newGam)
holds for $W=D$, so that the reduction procedure \(reduce) yields \(newGam)
for all $W$.  As a consequence, from \(Del) and the corresponding equation
$\Delta_\mutilde(f)=\tilde\Gamma(W)$, the translation sums satisfy \(same).

We next turn to the calculation of the translation sums associated with the 
particular family of functions
$f_n(\eta)=(\eta(0)-\rho_-)(\eta(n)-\rho_+)$, $n\ge1$.  
Let us write
 $$\Phi_n \equiv \Delta_\muhat(f_n)
   = \sum_{i=-\infty}^\infty\E{(\eta(i)-\rho_-)(\eta(i+n)-\rho_+)}_\muhat.
  \(Phin)  $$
 Since $f_n(\eta)-f_{n+1}(\eta)=(\eta(n+1)-\eta(n))(\rho_--\eta(0))
=(T^nF)(\eta)$, where $F(\eta)=(\eta(1)-\eta(0))(\rho_--\eta(-n))$,
\(surprise) implies that 
 $$ \Phi_n - \Phi_{n+1}
  =   (\rho_+ - \rho_-) [\rho_- - \E{\eta(-n)}_\muhat]
  =   (\rho_+ - \rho_-) [\E{\eta(n)}_\muhat - \rho_+],\(Phin2) $$
 where we have used the symmetry \(sym) of the profile.  Thus
 $$\Phi_n - \lim_{k\to\infty}\Phi_k
  =(\rho_+-\rho_-) \sum_{i=n}^\infty[\E{\eta(i)}_\muhat -\rho_+].\(Phint)$$
 But $\Phi_1$ can be evaluated from \(ALG1)--\(ALG4), \(D2), and
\(ALG6bis):
 $$\Phi_1
     =(\rho_+-\rho_-) \sum_{i=n}^\infty[\E{\eta(i)}_\muhat -\rho_+].
   ={p\rho_--q\rho_+\over p-q}-\rho_+\rho_-. \(Phi1)$$
   The right hand side of \(Phint) in the case $n=1$ can be evaluated from
formulas (4.1)--(4.5) of \[DLS] and shown also to be given by
\(Phi1);  thus $\lim_{k\to\infty}\Phi_k=0$ (as one would expect) and hence
 $$\Phi_n = (\rho_+ -\rho_-) \sum_{i=n}^\infty 
    [\E{\eta(i)}_\muhat - \rho_+ ]. \(Phin3)$$
 %

 To understand the consequences of equation \(Phin3) we recall \cite{DLS}
that in general the profile $\E{\eta(n)}_\muhat$ decays exponentially to its
asymptotic value: for $n\gg 1$,
 $$\E{\eta(n)}_\muhat -\rho_+ \simeq Cn^{\gamma}e^{-\alpha n},\(pasymp)$$
 where $C>0$ if $q/p<x^*$, $C=0$ (and in fact $\E{\eta(n)}_\muhat=\rho_+$ for
all $n>0$) if $q/p = x^*$, and $C<0$ if $q/p>x^*$, with $x^*=\rho_- (1-
\rho_+)/\rho_+ (1- \rho_-)$ (explicit values of $\alpha$ and $\gamma$ are
given in \[DLS]).  Thus \(Phin3) implies that, except on the line $q/p=x^*$,
$\Phi_n$ decreases exponentially to $0$ as $n \to \infty$, with
characteristic length $\alpha^{-1}$.  Since the $\Phi_n$ are intrinsic properties,
this gives an intrinsic characteristic size for the shock. 

A somewhat surprising aspect of the shock profile, found in \[DJLS] for $p=1$
and in \[DLS] for all $p$ satisfying $q/p<x^*$, is the ``overshoot'':
$\E{\eta(n)}_\muhat>\rho_+$ for $n\ge1$ (corresponding to $C>0$ in
\(pasymp)).  A natural question is whether this overshoot is a by-product of
the choice to view the shock from the second class particle, and might be
eliminated by the adoption of another viewpoint.  While we cannot answer this
completely, \(Phin3) implies that for $q/p<x^*$ there is no viewpoint from
which the shock measure is described by a product measure with density
increasing monotonically from $\rho_-$ to $\rho_+$, since such a measure
would lead to a negative $\Phi_n$.  Similarly, for $q/p\ne x^*$ there is no
point of view from which the shock measure would be a product measure with
density $\rho_+$ to the right of the origin and density $\rho_-$ to the left
of the origin, since for such a measure, all the $\Phi_n$ would vanish. 

 \newsection 6. Shocks in other ASEP models

Several different models, with ASEP dynamics but with specific initial
conditions, boundary conditions or minor modifications, have been shown to
give rise to shocks:
 \medskip{\parindent30pt
 \item{1.} An infinite system with, as initial condition, a product
measure with density $\rho_+$ to the right and $\rho_-$ to
the left of the origin.  This is the case discussed in the introduction and
in \[FKS,F2,DLS].
 \smallskip
 \item{2.} A system with ring geometry and a blockage bond, at which the jump
rate is less than that at other bonds in the system \[JL].
 \smallskip
 \item{3.} A system with ring geometry and a slow second class particle
\[D,M].
 \smallskip
  \item{4.} A system with open boundary conditions on its first order
transition line, discussed below and in \[DEHP].
 \medskip}
 \noindent
 Note that in examples 2 and 3 the shock in the ring geometry is caused by
the blocking bond or particle but is located far from it, in a region where
the usual ASEP dynamics holds.  We believe that, when the size of the system
goes to infinity, all of these different models lead in fact to shocks which
are shift equivalent, i.e., that one may choose for each shock a viewpoint
such that, seen from this viewpoint, all the shocks are described by the same
ramp measure.  (The relation of shocks observed in similar but more
complicated systems \[AHR] to those of the ASEP remains to be investigated.)
This equivalence is not obvious {\it a priori} but can be verified, at
least in cases where exact expressions permit exact computations, by showing
that all the translation sums $\Delta(f)$ are identical to those found in
Section~4. 

We illustrate this by computing the translation sums for the ASEP with open
boundary conditions, in the totally asymmetric ($p=1$) case.  This is a
system with $N$ sites, in which particles enter the system at site 1 with
rate $\alpha$ and escape at site $N$ with rate $\beta$.  In \[DEHP], it was
shown that the steady state of this system can be fully described by an
algebra rather similar to the one discussed in Section~4 (the weight of a
configuration can be written as the matrix element of a matrix product where a
matrix $D$ is used when a site is occupied and a matrix $E$ when the site is
empty).  Using this algebra, in principle, any expectation can be calculated
for systems of any size and for any values of the parameters $\alpha$ and
$\beta$.  In particular, it was shown that if $\alpha<\min\{\beta,1/2\}$ the
system is in a low density phase, described in the bulk by a product measure
with density $\alpha$, while if $\beta<\min\{\alpha,1/2\}$ the system is in a
similar high density phase, with bulk density $1-\beta$. 

On the line $\alpha=\beta < 1/2$ there is a first order phase transition
between these two phases.  When the system is on this phase coexistence line
one can show \[DEHP] that for large $N$ and for $i$ far from the boundaries
($i\gg1$ and $N-i\gg1$) the profile is linear in $i/N$,
 $$\E{\eta(i)}
  = {i(1-\alpha ) + (N-i)\alpha\over N} + O\left(1\over N \right),\(oprof)$$
 and by a similar computation, not given in \[DEHP], 
that nearest neighbor two point correlations behave in the same way:
 $$\langle \eta(i) \eta(i+1) \rangle 
  = {i (1- \alpha )^2 + (N-i) \alpha^2\over N} 
   + O\left(1\over N\right).\(ocorr)$$
 The interpretation of this behavior is that there is a shock in the
system, separating a region of low density $\alpha$ to its left and of high
density $1 - \alpha$ to its right, and that the linear dependence on $i/N$
arises because the shock is equally likely to be anywhere in the system
(apart from some corrections of order $1/N$ due to boundaries).  It seems
natural to suppose as well that the structure of the shock is independent
of its location, so that if we determine the shock position in the
configuration $\eta$ by some function $\ell(\eta)$ as described in
Section~2, and write $\mu_m$ for the system measure conditioned on
$\ell(\eta)=m$, then the measures $\mu_m$ are just translates of one
another, $\mu_m=T^{m-m'}\mu_{m'}$, for $N$ large and $m$ far from the
boundaries.

If such an interpretation is correct, then the expectation of any function of
the occupation numbers near site $i$ should for large $N$ show linear
behavior in $i/N$ similar to that of \(oprof) and \(ocorr).  The leading
terms give no information about the shock structure, but it follows also from
the picture described above that there is a probability of order $1/N$ that
the shock is located near site $i$, so that to describe the shock properties
one must compute the terms of order $1/N$ in expressions like \(oprof) and
\(ocorr).  However, these order $1/N$ terms may contain not only
contributions coming from the events where the shock is in the neighborhood
of $i$, but also contributions from boundary effects, in particular from the
precise definition of the coordinate $i$. 

We now show that we may calculate these order $1/N$ terms and thus obtain
additional support for this picture.  If $f=f(\eta(i+1),\eta(i+2),... 
\eta(i+k))$ is a function depending on the occupation numbers of sites
between $i+1$ and $i+k$, with $k\ll N$, then it can be shown that for $i$ far
enough from the boundaries the expectation of $f$ takes the form
 $$\E{f} 
   = {i \E{f}_{\nu_{1-\alpha}} + (N+b-i-k)\E{f}_{\nu_\alpha} \over N+b} 
      + {\gamma(W) \over N+b} + o \left( 1 \over N \right)\;.\(largeN)$$
 Here $b$ is a constant of order 1, independent of $f$ and $i$, and $W$ is
the linear combination of words of length $k$ associated to $f$, as in
Section~4 (see \(corr)).  Equation \(largeN) defines $\gamma(W)$; the correct
choice of $b$, which may be regarded as representing the boundary effects,
guarantees that $\gamma(W)$ is independent of $i$.  Equation \(largeN) is
trivial for $f=1$ (corresponding to $k=0$) and leads to
 $$\gamma(1)=0.\(oalg1) $$
  For $f=\eta(i)$, \(largeN) may be verified from the asymptotics derived in
\[DEHP]; this calculation leads to explicit values for $b$ and of $\gamma(D)$,
but the values of these parameters are not needed for the calculation of the
translation sums.  For other words $W$ one can use the algebra of \[DEHP] to verify
\(largeN) by induction on the length of $W$, and also show that the
$\gamma(W)$ have the following properties:
 $$ \eqalignno{\gamma(W(D+E)) &= \gamma(W) + r_-(W) &\(oalg2)\cr
  \gamma((D+E)W) &= \gamma(W) + r_+(W) &\(oalg3)\cr
   \gamma(a_1 W_1+ a_2 W_2) &= a_1 \gamma(W_1) + a_2 \gamma(W_2)&\(oalg4) \cr
  \gamma(W_1 DE W_2) &
  = \alpha (1- \alpha) [ \gamma(W_1D W_2) + \gamma(W_1 E W_2)]&\(oalg5)\cr}$$
Here $r_+$ and $r_-$ are defined with the densities $\rho_+= 1 - \alpha$
and $\rho_- =\alpha$.  Thus the $\gamma(W)$ satisfy the same rules as
$\Gamma(W)$ of Section~4, and one can conclude (see the argument following
\(newGam)) that the general expression of these $\gamma(W)$ is given by
 $$\gamma(W) = \Gamma(W) + \gamma(D)
      { \mu_+(W) - \mu_-(W)\over  \rho_+ - \rho_- }\(change)$$

Now suppose that the function $f$ satisfies
$\E{f}_{\nu_\alpha}=\E{f}_{\nu_{1-\alpha}}=0$, so that \(largeN) becomes
 $$\E{f}={\gamma(W)\over N}+o(1/N).\(largeN2)$$
 This result has a simple interpretation in terms of the heuristic picture,
described above, that the steady state of the open system contains a shock
equally likely to be located at any site and with structure independent of
its location.  Thus in computing $\E{f}$ we are averaging the expected value
in the measure conditioned on the shock position being $m$, $\E{f}_{\mu_m}$,
over all $N$ possible shock positions.  Since the measures $\mu_m$ differ
only by translation this is, up to a factor $1/N$, just the calculation of
the translation sum 
$\Delta_{\mu_m}(f)$ for any fixed $m$ ($N$ must be large enough
so that $\mu_m$ is effectively an infinite volume measure).  Thus \(largeN2)
implies that $\Delta_{\mu_m}(f)=\gamma(W)$.  But \(change) implies that for
such a function $f$, $\gamma(W)=\Gamma(W)$, so that from \(Del),
$\Delta_{\mu_m}(f)=\Delta_\muhat(f)$: the translation sums 
for the shock in the
open system are the same as those calculated in Section~4.  Thus these two
shocks are shift equivalent.

 \newsection Acknowledgments

We thank P.  Ferrari, Y.  Sinai, and G.  Steiff for helpful discussions.
This work was supported in part by NSF grant DMS~95--04556, NSF grant
DMR~95--23266, and AFSOR grant 4--26435.  BD thanks the Institute for
Advanced Study and JLL and ERS the Institut des Hautes Etudes Scientifiques
for hospitality while portions of this work were carried out; BD and JLL
would also like to thank DIMACS and its supporting agencies, the NSF under
contract STC-91-19999 and the N. J. Commission on Science and Technology.


\references

 \smallskip\noindent
 \[Ru] D. Ruelle, {\it Statistical Mechanics: Rigorous Results} (Benjamin,
New York, 1969).

 \smallskip\noindent
 \[G] H.-O. Georgii, {\it Gibbs Measures and Phase Transitions} (Walter de
Gruyter, Berlin, 1988).

 \smallskip\noindent
 \[Dob] R. Dobrushin, Gibbs states describing coexistence of phases for a
three-dimensional Ising model, {\it Th. Prob. Appl.} {\bf 17}:582--600 (1972).

 \smallskip\noindent
 \[Gal] G. Gallavotti, The phase separation line in the two-dimensional Ising
model, {\it Commun. Math. Phys.} {\bf 27}:103--136 (1972).

 \smallskip\noindent
 \[MM] A. Messager and S. Miracle-Sole, Correlation functions and boundary
conditions in the Ising ferromagnet, J.~Stat.~Phys {\bf 17}:245--262 (1977).

 \smallskip\noindent
 \[BLP] J. Bricmont, J. L. Lebowitz, and C. E. Pfister, On the local
structure of the phase separation line in the two-dimensional Ising system,
{\it J. Stat. Phys.} {\bf 26}:313--332 (1981).

\smallskip\noindent
 \[Spi] F. Spitzer, Interaction of Markov processes, 
    {\it Advances in Math.} {\bf 5}:246-290 (1970).

\smallskip\noindent
 \[Ligg] T. M. Liggett,   {\it Interacting Particle Systems}
(Springer-Verlag, New York, 1985). See also references therein.

\smallskip\noindent
 \[R] H. Rost, Nonequilibrium behavior of many particle process:
density profiles and local equilibria, {\it Z. Wahrsch. Verw. Gebiete}
{\bf 58}:41--53,  (1981).

\smallskip\noindent
 \[AV] E. D. Andjel and M. E. Vares, Hydrodynamical equations for
  attractive particle systems on $\bbz$, {\it J. Stat.\ Phys.} 
  {\bf 47}:265--288 (1987).

\smallskip\noindent
 \[Raz] F. Razakhanlou, Hydrodynamic limit for attractive particle
systems on $\bbz^d$, {\it Commun. Math. Phys.} {\bf 140}:417--448 (1991).

\smallskip\noindent
 \[ABL] E. D. Andjel, M. Bramson, and T. M. Liggett, Shocks in the
        asymmetric exclusion process, {\it Probab. Theory Relat. Fields} 
        {\bf 78}:231--247 (1988).

\smallskip\noindent
 \[FF] P. Ferrari and L.~R.~G.~Fontes, Shock fluctuations in the
asymmetric simple exclusion process, {\it Probab. Theory Relat. Fields} {\bf
99}:305--319 (1994).

\smallskip\noindent
 \[FKS] P. Ferrari, C. Kipnis, and E. Saada, Microscopic
        structure of traveling waves in the asymmetric simple
        exclusion, {\it Ann. Probab.} {\bf 19}:226--244 (1991).

\smallskip\noindent
 \[F2] P.  Ferrari, Shock fluctuations in asymmetric simple
exclusion, {\it Probab. Theory Relat. Fields} {\bf 91}:81--101 (1992).

 \smallskip\noindent
 \[DJLS] B. Derrida, S. A. Janowsky, J. L. Lebowitz, and E. R. Speer,
Exact solution of the totally asymmetric simple exclusion process:
shock profiles, {\it J. Stat. Phys.} {\bf 73}:813--842 (1993).

 \smallskip\noindent
 \[DLS] B. Derrida, J. L. Lebowitz, and E. R. Speer, Shock profiles in
the asymmetric simple exclusion process in one dimension, {\it J. Stat Phys.}
{\bf 89}:135--167 (1997).

 \smallskip\noindent
 \[DGLS2] Work in preparation.

 \smallskip\noindent
 \[O] B. M. Gurevi\vaccent c and V. I. Oseledec, Gibbs distributions and
dissipativeness of $U$-diffeomorphisms, {\it Soviet Math. Dokl} {\bf
14}:570--573 (1973).

 \smallskip\noindent
 \[ATh] D.  J.  Aldous and H.  Thorisson, Shift-coupling, {\it Stoch.  Proc. 
Appl.} {\bf 44}:1--14 (1993). 

 \smallskip\noindent
 \[Th] H.  Thorisson, On time- and cycle-stationarity, {\it Stoch.  Proc. 
Appl.} {\bf 55}:183--209 (1995). 

 \smallskip\noindent
 \[G2] H.-O. Georgii, Orbit coupling, {\it Ann. Inst. Henri Poincar\'e} {\bf
33}:253--268 (1997). 

 \smallskip\noindent
 \[Vass] L. N. Vasershtein, Markov processes over denumerable products of
spaces, describing large systems of automata, {\it Problems Inform.
Transmission,} {\bf 5}:47--52 (1969).

\smallskip\noindent
 \[Sandow] S. Sandow, Partially asymmetric exclusion process with open
      boundaries, {\it Phys. Rev. E} {\bf 50}:2660--2667 (1994).

 \smallskip\noindent
 \[JL] S. Janowsky and J. L. Lebowitz, Finite size effects and shock
fluctuations in the asymmetric simple exclusion process, {\it Phys. Rev. A}
{\bf 45}, 618--625 (1992), and Exact results for the asymmetric simple
exclusion process with a blockage, {\it J. Stat Phys.}
{\bf 77}:35--51 (1994).

 \smallskip\noindent
 \[D] B. Derrida, Systems out of equilibrium: some exactly solved models,
in {\it STATPHYS: the 19th IUPAP International Conference on Statistical
Physics},  ed. H. Bailin (World Scientific, Singapore, 1996).

\smallskip\noindent
 \[M] K. Mallick,  Shocks in the asymmetric simple exclusion model with
an impurity, {\it J. Phys. A} {\bf29}:5375--5386 (1996).

 \smallskip\noindent
 \[DEHP] B.  Derrida, M.  R.  Evans, V.  Hakim, and V.  Pasquier,
        An exact solution of a 1D asymmetric exclusion model using a
        matrix formulation, {\it J. Phys. A} {\bf 26}:1493--1517 (1993).

 \smallskip\noindent
 \[AHR] P. F. Arndt, T. Heinzel, and V. Rittenberg, 
First-order phase transitions in one-dimensional steady states, preprint
SISSA Ref. 39/37/EP, cond-mat/9706114.

\vfill\eject

\noindent 
{\bf Figure captions}

 \bigskip\noindent
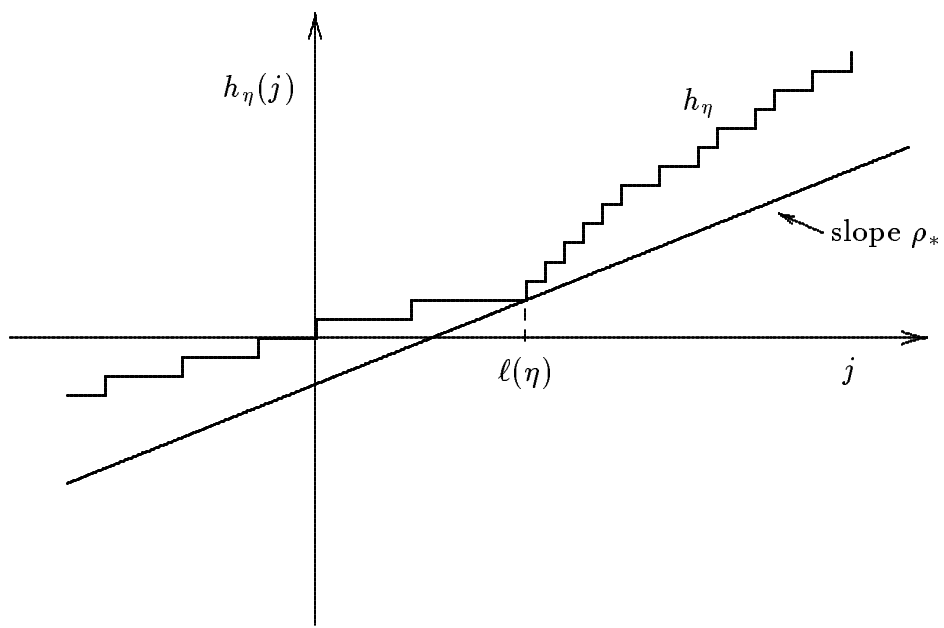
 {\bf Figure 1.} Construction of the intrinsic shock location function
$\ell(\eta)$. 
 \medskip\noindent
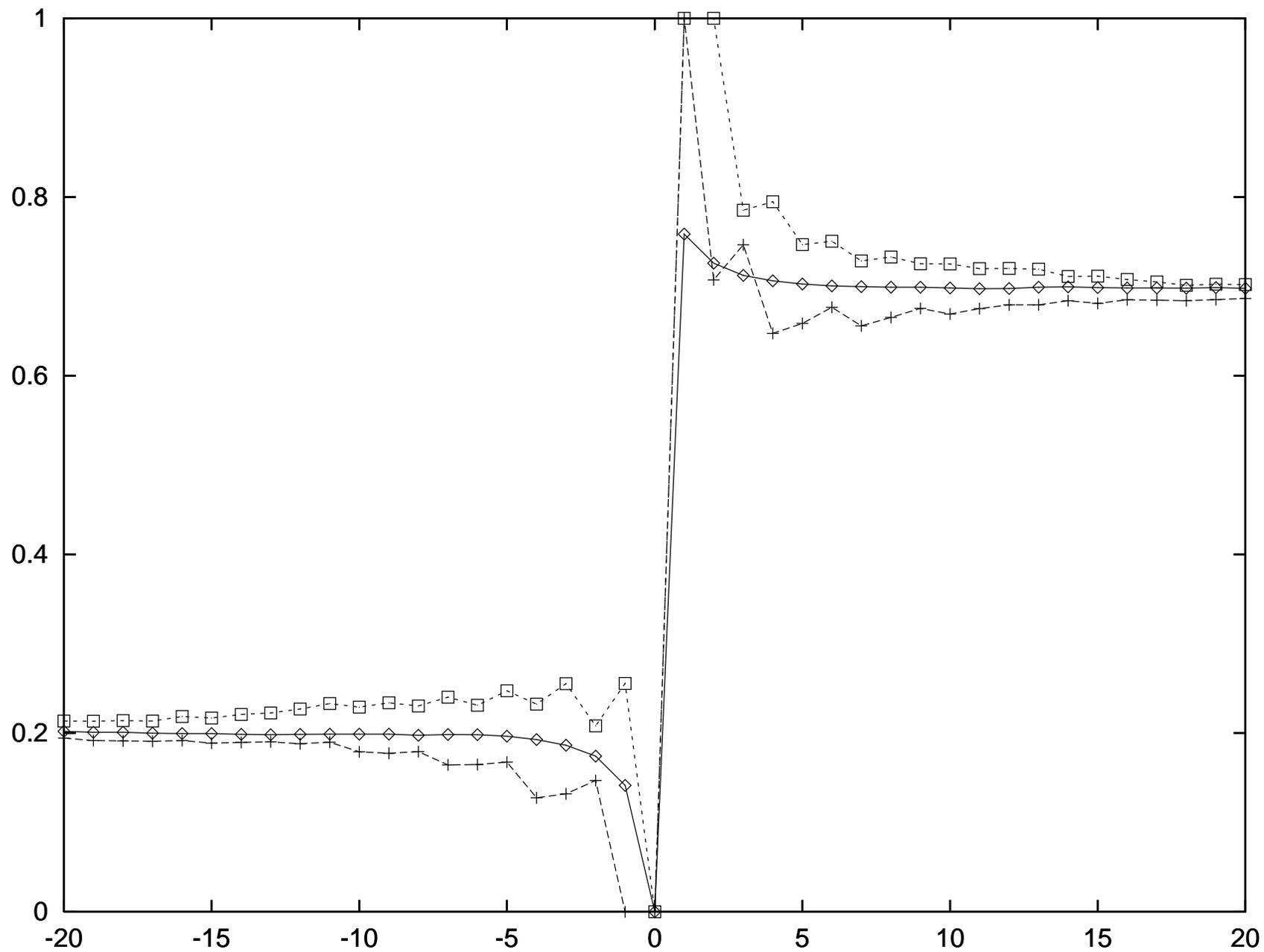
 {\bf Figure 2.}  Shock profiles for $p=1$, $\rho_+=0.7$,
$\rho_-=0.2$, from three different viewpoints: (i)~the second class particle
(diamonds, solid line), (ii)~$\ell$ defined with
$\rho_*=\pi^{-1}\rho_+ + (1-\pi^{-1})\rho_-$ (plusses, dashed line),
(iii)~$\ell$ defined with $\rho_*=(1-\pi^{-1})\rho_+ + \pi^{-1}\rho_-$
(squares, dotted line).

\bye